# DeepHIPS: A novel Deep Learning based Hippocampus Subfield Segmentation method

José V. Manjón[1], José E. Romero[1] and Pierrick Coupé[2,3]

[1]Instituto de Aplicaciones de las Tecnologías de la Información y de las Comunicaciones Avanzadas (ITACA), Universitat Politècnica de València, Camino de Vera s/n, 46022 Valencia, España.
[2] Univ. Bordeaux, LaBRI, UMR 5800, PICTURA, F-33400 Talence, France.
[3] CNRS, LaBRI, UMR 5800, PICTURA, F-33400 Talence, France.

**Abstract.** The automatic assessment of hippocampus volume is an important tool in the study of several neurodegenerative diseases such as Alzheimer's disease. Specifically, the measurement of hippocampus subfields properties is of great interest since it can show earlier pathological changes in the brain. However, segmentation of these subfields is very difficult due to their complex structure and for the need of high-resolution magnetic resonance images manually labeled. In this work, we present a novel pipeline for automatic hippocampus subfield segmentation based on a deeply supervised convolutional neural network. Results of the proposed method are shown for two available hippocampus subfield delineation protocols. The method has been compared to other state-of-the-art methods showing improved results in terms of accuracy and execution time.

# 1. Introduction

The hippocampus (HC) is a bilateral brain structure located in the medial temporal lobe at both sides of the brainstem near to the cerebellum. HC is involved in many brain functions such as memory and spatial reasoning (Milner et al., 1958; Schmajuk 1990). It plays also an important role in many neurodegenerative diseases such as Alzheimer's disease (AD) (Braak et al., 1991). Furthermore, hippocampus volume estimation is considered a valuable tool for follow-up and treatment adjustment (Jack et al., 2000; Jack et al., 2005; Dickerson and Sperling, 2005).

In the last years, many HC segmentation methods have been proposed (Barnes et al., 2008; Collins el al., 2010; Coupe et al., 2011). Most of them, were restricted to consider the hippocampus as a single structure (Chupin et al., 2009) due image resolution limitations. However, it is well known that, for example, AD affects the different HC subfields at different moments during the disease progression (Braak et al., 1991; Hett el al., 2019). Thus, automatic and accurate HC subfield segmentation methods would be really important to obtain early biomarkers of the disease.

Currently, advances in modern MR sequences allow acquiring high-resolution images making possible to divide the hippocampus into its constituent parts. In the last years, several delineation protocols have been proposed (some of these protocols have been used to create manually labeled MRI datasets). However, there is still little consensus between the different HC subfield protocols as shown in (Yushkevich et al., 2015a) where 21 delineation protocols were compared. For example, in 2013, Winterburn presented a new in-vivo high-resolution atlas (Winterburn et al., 2013) to divide the hippocampus in five different sub-regions: CA1, CA2-3, CA4/DG, Stratum and Subiculum. Later, in 2015, Kulaga-Yoskovitz developed another segmentation protocol (Kulaga-Yoskovitz et al., 2015) consisting of three structures: CA1-3, CA4/DG and Subiculum

Several automatic methods for HC subfield segmentation have been developed in the last years (Van Leemput et al., 2009, Pipitone et al., 2014, Iglesias et al., 2015). One of the most well-known methods for HC subfield segmentation is named ASHS (Yushkevich et al., 2015b) that uses a multi-atlas approach combined with a similarity-weighted voting and a boosting-based error correction. Unfortunately, this method takes several hours to produce a segmentation due to the exhaustive use of non-linear registrations. More recently, we proposed a method named HIPS (Romero et al., 2017) that obtained state-of-the-art results in two different delineation protocols (Winterburn and Kulaga-Yoskovitz) with relatively low processing times thanks to the use a fast multiatlas label fusion method called OPAL (Giraud et al., 2016). Although these methods have promising results, their automatic measurements are not close enough to manual tracings in some cases (Peixoto-Santos et al., 2018).

Recently, due to the expansion of deep learning in medical imaging, novel methods based on this technology have been proposed to further improve the accuracy of HC segmentation. For full hippocampus segmentation many methods based on convolutional neural networks (CNN) have been already proposed (Chen et al., 2017; Cao et al., 2018; Thyreau et al. 2018; Ataloglou et al. 2019). Recently, deep learning based methods has been also proposed for hippocampus subfield segmentation. For example, UGNET has been proposed (Shi et al., 2019) using an adversarial training approach and also variants of the famous UNET architecture (Ronneberger et al., 2015) such as the Dilated Dense UNET (Zhu et al.,2019) have been proposed. However, one of the major problems of supervised deep learning methods is their hunger for training data to be able to generalize on unseen data.

In this paper, we propose a novel deep-learning based segmentation method that takes benefit of a problem specific preprocessing that locates the data in a canonical geometrical and intensity space therefore simplifying the segmentation problem and thus reducing the need for lots of manually labeled data. The proposed method has been validated using two hippocampus subfield segmentation protocols with publically available datasets.

## 2. Material and methods

### 2.1. Training data

In this work, we have used two different datasets including two manual labeling hippocampus subfield segmentation protocols, both with high-resolution (HR) T1w and T2w MR images (figure 1). Details of these datasets are given below:

*Kulaga-Yoskovitz dataset*

This dataset includes 25 subjects from a public repository (http://www.nitrc.org/projects/mni-hisub25) (31 ± 7 yrs, 12 males, 13 females) with manually segmented labels dividing the HC in three parts (CA1-3, DG-CA4 and Subiculum). MR data from each subject consist of an isotropic 3D-MPRAGE T1-weighted (0.6 mm$^3$) and anisotropic 2D T2-weighted TSE images (0.4×0.4×2 mm$^3$). Images underwent automated correction for intensity non-uniformity, intensity standardization and were linearly registered to the MNI152 space. T1w and T2w images were resampled to a resolution of 0.4 mm$^3$. To reduce interpolation artifacts, the T2w data was upsampled using a non-local super-resolution method (Manjón et al., 2010a). For more details about the labeling protocol see the original paper (Kulaga-Yoskovitz et al., 2015).

*Winterburn dataset*

This dataset contains 5 subjects with 0.3x0.3x0.3 mm$^3$ high resolution T1-weighted and T2-weighted images obtained by 2x interpolation of 0.6x0.6x0.6 mm$^3$ acquisitions and their corresponding manual segmentations. The HR images are publicly available at the CoBrALab website (http://cobralab.ca/atlases). These MR images were taken from 5 healthy volunteers (2 males, 3 females, aged 29–57). High-resolution T1-weighted images were acquired using the 3D inversion-prepared fast spoiled gradient-recalled echo acquisition (TE/TR=4.3 ms/9.2 ms, TI=650 ms, α=8°, 2-NEX and isotropic resolution of 0.6 mm$^3$). High-resolution T2-weighted images were acquired using the 3D fast spin echo acquisition, FSE-CUBE (TE/TR=95.3 ms/2500 ms, ETL=100 ms, 2NEX, and isotropic resolution of 0.6 mm$^3$). Reconstruction filters, ZIPX2 and ZIP512, were also used resulting in a final isotropic 0.3 mm$^3$ dimension voxels. The hippocampi and each of their subfields were segmented manually by an expert rater including 5 labels (CA1, CA2/3, CA4/DG, (SR/SL/SM), and subiculum). For more details about the labeling protocol see the original paper (Winterburn et al., 2013).

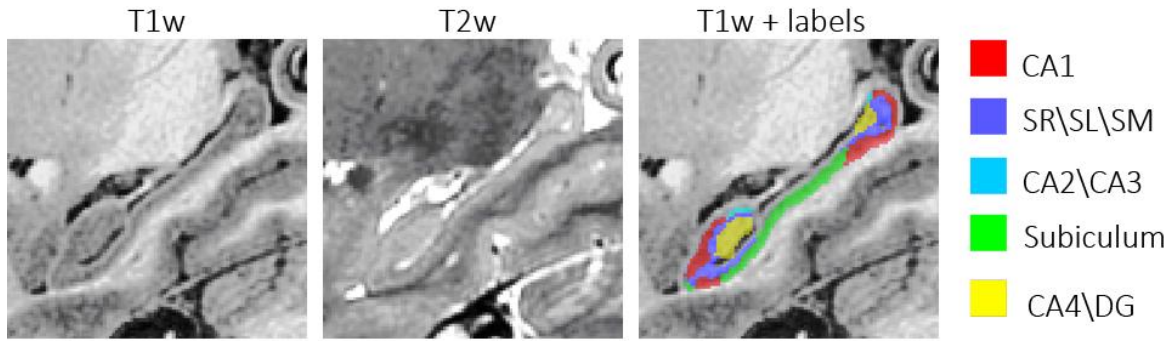
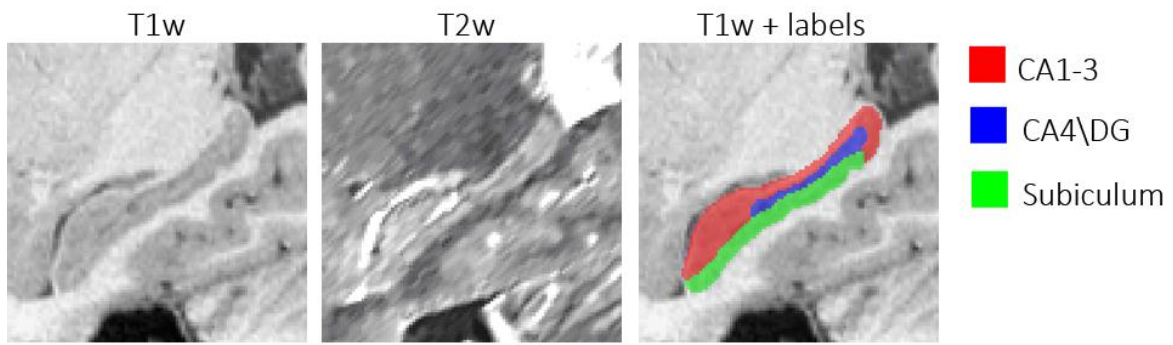

*Figure 1*: Examples from Winterburn and Kulaga-Yoskovitz datasets showing T1w, T2w and manual segmentations.

**2.2. Image preprocessing**

The images were preprocessed using the following steps: 1) Denoising using the Spatially Adaptive Non-Local Means Filter (Manjón et al., 2010b), 2) Intensity inhomogeneity correction using the N4 bias field correction (Tustison et al., 2010), 3) Registration to the Montreal Neurological Institute (MNI) space by applying the Advanced Normalization Tools (ANTs) package (Avants et al., 2009). This registration was estimated using the T1w MNI152 template (at 0.5 mm$^3$ resolution) and the T1w images, and applied to both T1w and T2w images (a rigid transformation from T2w to T1w was previously estimated and later concatenated with T1w transformation to perform a single interpolation step when registering both T1w and T2w images). 4) Cropping: To reduce the memory requirements and the computational cost, the images were cropped around HC area, 5) Finally, the cropped images were intensity normalized by subtracting the image mean and dividing by its standard deviation.

## 2.3. Proposed method

Our proposed method is based on a variant of the well-known UNET architecture (Ronneberger et al., 2015). The proposed UNET has 4 resolution levels (from 0.5 to 4 mm). We used three blocks of BatchNormalization, 3D convolution (kernel size of 3x3x3 voxels) plus ReLU layers for each resolution level. We also used dropout layers (with 0.5 rate) in the encoding part of the UNET to minimize overfitting problems.

The input of the network consists of a tensor with two channels (T1 and T2 images). The first resolution level has 64 filters and the next levels multiply by 2 this number to compensate the loss of spatial resolution. Similarly, the number of filters is reduced by 2 in the ascending path of the encoder at each resolution level. The output is also a tensor of *nc* channels represent the probabilities of each subfield and the background.

We also used a modified version of deep supervision (Dou et al., 2017) approach that helps to train very deep networks by producing segmentations at different resolution levels. Deep supervision has been shown to not only counteract the adverse effects of gradient vanishing but also to speed up convergence and produce highly accurate results even with limited data. The main difference of our implementation compared to Dou et al., is that we use upsampled low-resolution outputs also as inputs of the next level of the decoder (concatenated with the upsampled features and the encoder shortcut) to help in the next resolution level (only for 1 and 2 mm resolution levels). The resulting network has 56 layers and 35,085,580 trainable parameters. In figure 2 the scheme of the proposed network is shown.

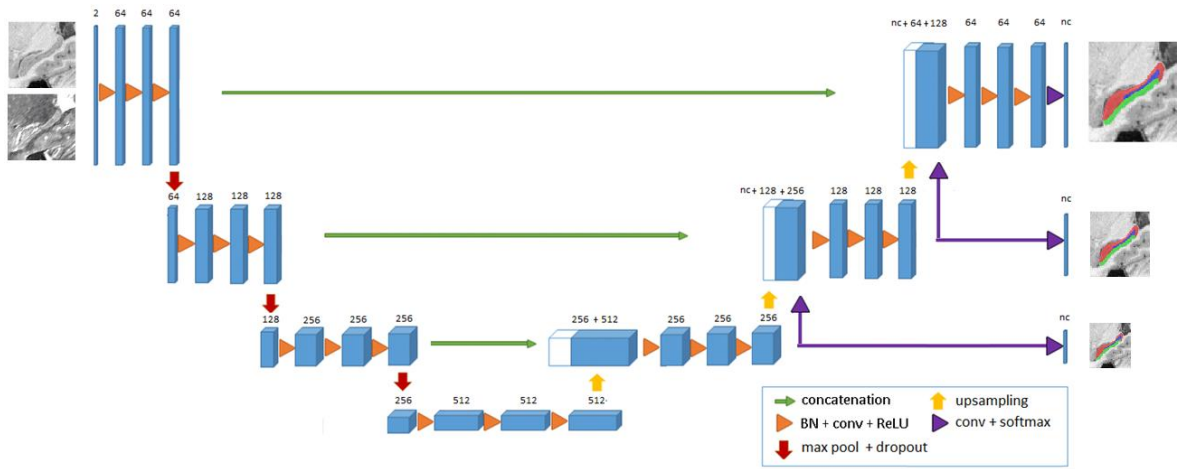

*Figure 2.* Scheme of the proposed Deep Supervised UNET CNN.

The loss function plays a major role in the training process and an exhaustive search of the most suitable function for the proposed architecture and problem has to be done. One of the most common loss functions for classification is the categorical cross entropy. However, for segmentation purposes it is common to use the dice loss (DL) (Milletari et al., 2016) as it directly optimizes the segmentation metric most commonly used and it is more robust to the class imbalance problem (1). Recently, a Generalized Dice Loss (GDL) (Sudre et al., 2017) was proposed to deal with the well-known dependency of the dice index and the size of the labels (2). Inspired by GDL, we proposed in this paper the Generalized Jaccard Loss (GDL) (4) which is a variant of Jaccard loss (3) following the same idea to reduce label size dependency:

$$DL(p,t) = 1 - \frac{2}{NC}\sum_{c=1}^{NC}\frac{\sum_{i=1}^{N}p_{ci}t_{ci}}{\sum_{i=1}^{N}p_{ci}+t_{ci}} \quad (1)$$

$$GDL(p,t) = 1 - 2\frac{\sum_{c=1}^{NC}w_c\sum_{i=1}^{N}p_{ci}t_{ci}}{\sum_{c=1}^{NC}w_c\sum_{i=1}^{N}p_{ci}+t_{ci}} \quad (2)$$

$$JL(p,t) = 1 - \frac{1}{NC}\frac{\sum_{c=1}^{NC}\sum_{i=1}^{N}p_{ci}t_{ci}}{\sum_{c=1}^{NC}(\sum_{i=1}^{N}p_{ci}+t_{ci}-\sum_{i=1}^{N}p_{ci}t_{ci})} \quad (3)$$

$$GJL(p,t) = 1 - \frac{\sum_{c=1}^{NC}w_c\sum_{i=1}^{N}p_{ci}t_{ci}}{\sum_{c=1}^{NC}w_c(\sum_{i=1}^{N}p_{ci}+t_{ci}-\sum_{i=1}^{N}p_{ci}t_{ci})} \quad (4)$$

where $N$ is the number of voxels, $NC$ is the number of classes, $p$ is the predicted probability and $t$ is the true probability and $w_c = 1/(\sum_{i=1}^{N}t_{ci})^2$. Alternatively, we define the GDL* and GJL* losses for the specific case of not using the squared volume to normalize but just the volume, i.e. $w_c = 1/\sum_{i=1}^{N}t_{ci}$ .

As the size of our training datasets is small (specially for the Winterburn dataset) we used different approaches of data augmentation. We randomly smooth and sharpen the images to simulate different image quality data during the training. We expanded the Winterburn dataset with automatic segmentations of the kulaga-Yoskovitz dataset using the method HIPS. Note that to generate these segmentations only training data was used as library. Finally, we used also mixup (Zhang et al., 2017) as a data agnostic method for performing data augmentation.

Batch normalization is a highly effective manner to speed up the training process and to improve the results by minimizing the internal covariate shift. However, we realized that when used with small batch sizes it behaves sub-optimally at test time. The reason of this issue is that Batch Normalization layer behaves differently at training and test time. During training, the mean and standard deviation of the activation maps are computed for the whole batch using a moving average estimation to enforce stability during training. However, at test time the network does not processes any batch of data and therefore cannot estimate the mean and standard deviation of the batch, as a result, the network uses the historical mean and standard deviation stored during training. Unfortunately, when using small batch sizes (N=1 in our case) the stored values do not work very well. Nevertheless, if we run the network in training mode we force the network to use the current mean and standard deviation of the new case and the results are significantly improved. We call this, training time batch normalization (TTBN).

## 3. Experiments and results

In this section, the analysis of the different options of the proposed method and their results are presented. To evaluate the segmentation accuracy, we have used the DICE coefficient (Zijdenbos et al., 1994) measured in the linear MNI152 space. All experiments were performed using tensorflow 1.2.0 and keras 2.2.4 using Titan Xp Nvidia GPU with 12 GB RAM. To train the network, we used an Adam optimizer (Diederik et al., 2014) with default parameters during 200 epochs and we test different loss functions with multiscale loss weights (0.1, 0.2 and 0.7) for low, medium and high-resolution outputs respectively (see Fig. 2). A batch size of one was used in all our experiments.

Kulaga-Yoskovitz and Winterburn datasets were preprocessed as described in section 2.2. To increase the size of the training data, we put together left and right crops by left-right flipping the left crops to generate right oriented crops. This yield 50 right crops in Kulaga-Yoskovitz dataset and 10 right crops in Winterburn dataset. Since the size of both datasets is quite small we have used a K-fold cross validation strategy to increase the relevance of our findings. Specifically we used K=5 in both datasets. In Kulaga-Yoskovitz dataset this let each fold with 40 training images (5 of them for validation) and 10 test images. In Winterburn dataset each fold had 8 training images (2 of them for validation) and 2 test images.

### 3.1. Analysis of the proposed method

There are many factors that affect the performance of deep learning methods such as the architecture, the loss function, data augmentation strategies, etc. In this section we will present some experiments that show their effects on the proposed method. To perform these experiments, we selected one fold of both datasets and we trained it to evaluate the validation DICE.

The first option we tested was the loss function. We compared 6 different loss functions using the same exact network initialization. We repeated this experiment 5 times to robustly estimate the network performance for both datasets. In table 1 the validation DICE of each loss function is compared for both datasets. We also included the categorical cross entropy (CCE) in the comparison as it is a common loss used in segmentation/classification. As can be noted, CCE performed worse than dice loss which is a common loss function used in segmentation. Curiously, the GDL failed to train giving a really low dice. Changing the volume-based weight from quadratic to linear (GDL*) stabilized the training and provided better results but no better that dice loss. JL performed slightly worse than dice loss (specially in Winterburn dataset). The proposed GJL* was the best performing loss in both datasets and therefore was selected a loss function of the proposed method. GJL had the same issues than GDL (data not shown).

*Table 1.* Average DICE in Kulaga-Yoskovitz (first row) and Winterburn (second row) datasets.

| CCE | DL | GDL | GDL* | JL | GJL* |
|---|---|---|---|---|---|
| 0.8822±0.0067 | 0.9095±0.0013 | 0.3562±0.0039 | 0.9092±0.0038 | 0.9078±0.0014 | **0.9116±0.0014** |
| *0.6042±0.0155* | *0.6634 ±0.0147* | *0.1807±0.0148* | *0.6626±0.0056* | *0.6465±0.0086* | ***0.6704±0.0045*** |

To study the impact of the proposed architecture, we run 5 times the proposed network and we compared with the classic UNET without deep supervision and feedback. In table 2 the results of the comparison are shown. As can be noticed, the proposed architecture was able to improve the results in both datasets being the Winterburn dataset the one that had the higher benefit (probably due to the lower number of training images).

*Table 2.* Comparison of our proposed deep supervised UNET vs classic UNET 3D.

|  | **UNET 3D** | **Proposed UNET 3D** |
|---|---|---|
| *Kulaga-Yoskovitz* | 0.9083±0.0030 | **0.9116±0.0014** |
| *Winterburn* | 0.6148±0.0354 | *0.6704±0.0045* |

It is well-known that in deep learning the amount of training data plays a major role (probably the biggest) in the quality of the network results. Unfortunately, manually labeled cases of hippocampal subfields is a rare resource due to the difficulty of generating such data. Automatic data augmentation has been traditional used to artificially increase the number of training cases. This has been usually done applying random transformations on the available training data (rotation, scale, etc.). In this project, we have used a combination of different methods to augment the number of training cases. In the case of Kulaga-Yuskevitz, we randomly smooth and sharpen the cropped images to generate low and high-quality images to improve generalization capabilities of the network. We also used mixup (Zhang et al., 2017) to linearly combine inputs and outputs (alfa=0.3). Mixup is a data-agnostic data augmentation method that has been proven beneficial specially when using a small training dataset (Eaton-Rosen et al., 2018). In the case of Winterburn dataset, we used the same approach but in addition we increased the training dataset using automatic segmentations of the Kulaga-Yuskevitz dataset with the HIPS method (Romero et al., 2017) which is a patch-based multiatlas label fusion based method (using as atlases the training cases of the fold). In the table 3 the results of the proposed method with and without data augmentations are shown. As expected, data augmentation strategies helped to improve the results in both datasets. The improvement in Winterburn dataset was more important given the small size of the training set (N=6).

*Table 3.* Data augmentation results

|  | No data augmentation | Data augmentation |
|---|---|---|
| *Kulaga-Yoskovitz* | 0.9116±0.0014 | *0.9139±0.0012* |
| *Winterburn* | *0.6704±0.0045* | *0.7048±0.0048* |

A last experiment was performed to evaluate the effect of the TTBN technique. In table 4 the results of both datasets are shown. As can be noticed, TTBN help in both datasets being the improvement of Winterburn dataset relatively higher.

*Table 4.* Training Time Batch Normalization results.

|  | Standard prediction | TTBN prediction |
|---|---|---|
| *Kulaga-Yoskovitz* | *0.9139±0.0012* | *0.9160±0.0011* |
| *Winterburn* | *0.7048±0.0048* | *0.7116±0.0018* |

Finally, we trained the 5 folds including all discussed optimizations to obtain the final dice of the proposed method for each dataset for 200 epochs. We included the validation data into the training data to further increase the number available data for training. We estimated each structure dice, average dice among structures and whole hippocampus dice. In table 5 the k-fold (k=5) cross validation results for both datasets are shown.

*Table 5. Mean DICE and standard deviation for each structure segmentation over the Kulaga-Yoskovitz and Winterburn datasets.*

| Structure\Protocol | Kulaga-Yoskovitz | Structure\Protocol | Winterburn |
|---|---|---|---|
| *Average* | 0.9037±0.0129 | *Average* | 0.7418±0.0188 |
| *CA1-3* | 0.9245±0.0106 | *CA1* | 0.7805±0.0170 |
| *CA4\DG* | 0.8887±0.0237 | *CA2\CA3* | 0.6686±0.0436 |
| *Subiculum* | 0.8980±0.0155 | *CA4\DG* | 0.8096±0.0301 |
|  |  | *SR\SL\SM* | 0.7066±0.0197 |
|  |  | *Subiculum* | 0.7439±0.0338 |
| *Hippocampus* | 0.9618±0.0051 | *Hippocampus* | 0.9123±0.0106 |

### 3.2. Method comparison

The proposed method was compared with state-of-the-art related methods. Specifically, for the Kulaga-Yoskovitz dataset we compared with HIPS method (Romero et al., 2017) and a recent deep learning-based method named ResDUnet dedicated to hippocampus subfields segmentation (Zhu et al., 2019). In both cases we used publised results in their papers for the comparison. In table 6 we show the results of the comparison. We included also the inter and intra-rater accuracy for comparison purposes. As can be noticed, the proposed method outperformed previous state-of-the-art methods. It is also worth to note that the proposed method improved the inter-rater accuracy and got very close to the intra-rater accuracy.

*Table 6: Mean DICE and standard deviation for each structure segmentation over the Kulaga-Yoskovitz dataset. Best results in bold.*

| Structure | HIPS | ResDUnet | Proposed | *Inter-rater* | *Intra-rater* |
|---|---|---|---|---|---|
| *Average* | 0.8879 | 0.8960 | **0.9037** | 0.8833 | 0.9113 |
| *CA1-3* | 0.9158±0.0150 | 0.9200±0.0110 | **0.9245±0.0106** | 0.8760 ± 0.048 | 0.9290 ± 0.010 |
| *CA4\DG* | 0.8863±0.0340 | 0.8790±0.0200 | **0.8887±0.0237** | 0.9030 ± 0.036 | 0.9000 ± 0.019 |
| *Subiculum* | 0.8616±0.0210 | 0.8880±0.0160 | **0.8980±0.0155** | 0.8710 ± 0.053 | 0.9050 ± 0.016 |
| *Hippocampus* | 0.9595 | ---- | **0.9618** | ---- | ---- |

*For the Winterburn dataset,* we compared with HIPS method (Romero et al., 2017) that represents the state of the art in this dataset. In table 7, we show the results of the comparison. We included also the intra-rater accuracy for comparison purposes. As can be noticed the proposed method outperformed HIPS method by a large margin and got very close to the intra-rater accuracy.

*Table 7: Mean DICE in the MNI space and standard deviation for each structure segmentation using high resolution T1w, T2w and Multispectral respectively over the Winterburn dataset. Best results in bold.*

| Structure | HIPS | Proposed | Intra-rater |
|---|---|---|---|
| *Average* | 0.7158 | **0.7418** | 0.742 |
| *CA1* | 0.7762±0.0251 | **0.7805±0.0170** | 0.780 |
| *CA2\CA3* | 0.6179±0.0630 | **0.6686±0.0436** | 0.640 |
| *CA4\DG* | 0.7750±0.0307 | **0.8096±0.0301** | 0.830 |
| *SR\SL\SM* | 0.7018±0.0191 | **0.7066±0.0197** | 0.710 |
| *Subiculum* | 0.7082±0.0597 | **0.7439±0.0338** | 0.750 |
| *Hippocampus* | 0.9111 | **0.9123** | 0.910 |

Regarding to the execution time, the proposed network takes around 1 second to segment a new case. The whole Deep HIPS pipeline (including preprocessing) takes around 2 minutes while HIPS method takes around 20 minutes.

## 4. Discussion

In this paper, we have presented a new deep learning-based method for HR hippocampus subfield segmentation that we called DeepHIPS. We have validated the proposed method using 2 publically available datasets (Winterburn and Kulaga-Yoskovitz).

Our proposed method first preprocess the HR T1 and T2 images to improve their quality and to locate them on a standard space (MNI152) to finally crop the region of interest to process. From the architecture point of view, our model is a 3D UNET variant that uses deep supervision and low-resolution feedback to make easier the training process. We found that this variant worked better that the classic UNET (specially for the Winterburn dataset).

We have proposed a novel loss function (GJL) based on the Jaccard similarity index that enables to improve the accuracy of the network borrowing ideas from a modified version of the GDL (i.e. using linear volume weights instead of quadratic). We further improve the results using classical data augmentation techniques such as image mirroring and intensity transformation and more modern ones such as mixup.

Finally, we improved the results of the network at test time by running Batch normalization layers in training mode instead of test mode. We found that when using small batch sizes (N=1 in our case) batch normalization layers did behave properly due to the use of the stored mean and standard deviation during training. Using current sample statistics systematically improved the results in all our experiments despite the simplicity of the approach. We called this Training Time Bach Normalization (TTBN).

We compared the results of the proposed method with state-of-the-art methods in two datasets. In the Kulaga-Yoskovitz dataset we compared with HIPS method and a recent deep learning-based method named ResDUnet. The proposed method improved the results of both methods for all subfields and got closer to the intra-rater accuracy which can be considered as the upper bound of the method. For the Winterburn dataset, we compared with HIPS method and again, the proposed method improved the results for all subfields and the overall accuracy got very close to the intra-rater accuracy.

We are aware that the training libraries of the proposed method are quite small to ensure a good generalization (especially in the case of Winterburn) and our future efforts will be directed to increase the size of these libraries by manually labeling new cases and using semi-supervised approaches to automatically extend the training dataset size.

From an efficiency point of view the proposed method is not only more accurate but also more efficient than previous state of the art (HIPS) reducing by a factor 10 the total execution time.

## 5. Conclusion

In this work, we have presented a new method for HR hippocampus subfield segmentation based on a deep learning approach and we have validated it with two publically available datasets (Winterburn and Kulaga-Yoskovitz) showing competitive results in both accuracy and efficiency. We plan to make fully accessible the DeepHIPS pipeline through the new release of our online image analysis service volbrain (http://volbrain.upv.es) so researchers around the world can use our pipeline without requiring complex pipeline installations or the use of expensive hardware (GPUs, etc).

## Acknowledgements

This research was supported by the Spanish DPI2017-87743-R grant from the Ministerio de Economia, Industria y Competitividad of Spain. This study has been also carried out with financial support from the French State, managed by the French National Research Agency (ANR) in the frame of the Investments for the future Program IdEx Bordeaux (ANR-10-IDEX-03-02, HL-MRI Project) and Cluster of excellence CPU and TRAIL (HR-DTI ANR-10-LABX-57). The authors gratefully acknowledge the support of NVIDIA Corporation with their donation of the TITAN X GPU used in this research.